\documentclass[twoside]{report}
\usepackage{iwsm}
\usepackage{graphicx}
\usepackage{amsmath, amssymb}
\usepackage{booktabs}
\usepackage{csvsimple}
\usepackage{siunitx}
\usepackage{float}
\usepackage{url}
\usepackage{natbib}
\usepackage{subfig}  % For subfigures
\begin{document}
\title{Bayesian Fractional Polynomials for Optimal Dosage Estimation with Fish Nutrition Applications}
\titlerunning{Bayesian Fractional Polynomial Modeling for Optimal Dosage}
\author{Aliaksandr Hubin\inst{1}, Åshild Krogdahl\inst{1}, Guro Løkka\inst{1}, Trond M. Kortner\inst{1}}
\authorrunning{Hubin}
\institute{NMBU, Norway}
\email{aliaksandr.hubin@nmbu.no}
\abstract{
The problem of optimal dosage estimation arises in diverse scientific domains, from pharmacology and toxicology to aquaculture and environmental studies. Statistical modeling of nonlinear dose–response relationships is essential to quantify biological effects and determine response-optimal levels. This paper introduces a flexible Bayesian fractional polynomial (BFP) framework for modeling such relationships, allowing for model uncertainty quantification and robust prediction through Bayesian model averaging. Extensive simulation results demonstrate that the proposed BFP approach yields accurate estimation of optimal dose levels, outperforming benchmarks significantly. The approach is demonstrated on real data from fish nutrient requirement experiments. 
}
\keywords{Bayesian fractional polynomials; model averaging; optimal dosage}
\maketitle
\section{Introduction} Accurately estimating the optimal level of an exposure or dosage is a central question in many applied sciences. Whether the goal is to find a therapeutic dose that maximizes patient benefit, a nutrient level that optimizes animal growth, or a contaminant concentration that minimizes risk, the relationship between the dose and response is rarely linear. Instead, it often displays nonlinear, saturating, or multimodal behavior, and the underlying functional form is typically unknown.

Traditional regression approaches such as polynomial fitting, locally estimated scatterplot smoothing (LOESS), and generalized additive models (GAMs) can represent nonlinearities, but they do not formally account for model (functional) uncertainty. Gaussian processes (GPs) provide model uncertainty quantification through their posterior distributions over functions but rely on a fixed kernel structure, limiting flexibility. In contrast, the Bayesian fractional polynomial (BFP) approach incorporates flexible basis functions within a probabilistic model that integrates rather broad functional uncertainty directly into inference. By performing Bayesian model averaging across combinations of candidate transformations, the method yields the full posterior distributions for dose-response relationship.

\section{Model Formulation}
The BFP model is formulated in the exponential family framework with:
\[
h(\mathbb{E}[y_i]) = \eta_i = \beta_0 + \sum_{k=1}^{K} \gamma_k \beta_k f_k(x_i), \quad i \in \{1,\cdots,n\},
\]
where \(n\) is the sample size, \( y_i \) is the response for dose \( x_i \), \( h(\cdot) \) is a link function, \( f_k(\cdot) \) are fractional polynomial transformations, and \( \gamma_k \in \{0, 1\} \) are inclusion indicators that determine whether a particular transformation enters the model. Each transformation is either \( f_k(x) = x^{p_k}\) or \( f_{k+8}(x) = x^{p_{k}} \ln x \), where the polynomial transformations are from a set \(\left\{p_k\right\}_{k=1}^{8} = \{-2, -1, -0.5, 0, 0.5, 1, 2, 3\}\), with a convention \(x^0 = \ln x\). Thus,  \(K = 16\).
Regression coefficients are assigned g-priors conditional on inclusion:
\[
\boldsymbol{\beta} | {\boldsymbol{\gamma}}, \sigma^2  \sim \mathcal{N}\big(0, g\cdot\sigma^2 (X_{\boldsymbol{\gamma}}^{\top} X_{\boldsymbol{\gamma}})^{-1}\big),
\]
where \( g \) is a global shrinkage parameter (e.g., set via empirical Bayes), and \( X_{\boldsymbol{\gamma}} \) is the design matrix for the active predictors defined by \({\boldsymbol{\gamma}}\). Lastly, \(\gamma_k\) have independent Bernoulli priors 
\[
\gamma_k\sim\text{Bernoulli}(0.5), \quad k \in \{1,\cdots,K\}.
\]

Since we only have one covariate, the posterior inference and model space exploration are carried out using mode-jumping MCMC (MJMCMC) \citep{Hubin2018mode} implemented in the \texttt{FBMS} R package \citep{frommlet2025fbms} and avoids the more computationally heavy genetically modified MJMCMC required for high dimensional BFP \citep{hubin2023fractional}. 

We define a model space $\mathcal{M}$ consisting of all $2^K$ combinations of fractional polynomial transformations, where each model $\mathrm{m} \in \mathcal{M}$ is defined by the inclusion vector $\gamma^{(\mathrm{m})} = (\gamma_1, \dots, \gamma_K)^\top$. 
MJMCMC algorithm efficiently navigates the high-dimensional model space $\mathcal{M}$, ensuring good mixing and after a finite number of iterations explores a subspace $\mathcal{M}^{\star}$. Then, posterior model probability for a model $\mathrm{m}$ on a subspace $\mathcal{M}^{\star}$ is obtained by Bayes' theorem:
\[
p_{\mathcal{M}^{\star}}(\mathrm{m} | \mathcal{D}) = \frac{p(\mathcal{D} | \mathrm{m}) p(\mathrm{m})}{\sum_{\mathrm{m}' \in \mathcal{M}^{\star}} p(\mathcal{D} | \mathrm{m}') p(\mathrm{m}')},
\]
where \(p(\mathcal{D} | \mathrm{m})\) is the marginal likelihood and \(p(\mathrm{m})\) is the prior probability of model \(\mathrm{m}\) and $\mathcal{D}$ is the training data sample. The marginal likelihood is computed analytically for the BFP framework with Gaussian responses and with Laplace approximations otherwise \citep{frommlet2025fbms}.

Posterior summaries are computed by averaging over the explored subspace \(\mathcal{M}^{\star}\). For a quantity of interest \(\Delta\), such as the optimal dose or a predicted response at some \(x\), the posterior distribution is given by
\[
p_{\mathcal{M}^{\star}}(\Delta |\mathcal{D}) = \sum_{\mathrm{m} \in \mathcal{M}^{\star}} p(\Delta |\mathcal{D}, \mathrm{m})\, p_{\mathcal{M}^{\star}}(\mathrm{m} |\mathcal{D}),
\]
where \(\mathrm{m}\) denotes a candidate fractional polynomial model and \(p(\Delta |\mathcal{D}, \mathrm{m})\) is the marginal posterior probability of $\Delta$ within model $\mathrm{m}$. Correspondingly, point summaries can be derived from the model-averaged posterior, including the posterior mean, median, and credible intervals. Futher, the highest posterior mass model (HPM), defined as
\(
\mathrm{m}_{\mathrm{HPM}} = \arg\max_{\mathrm{m} \in \mathcal{M}^{\star}} p(\mathrm{m} |\mathcal{D})
\), for which one can obtain $p(\Delta |\mathcal{D}, \mathrm{m}_{\mathrm{HPM}})$ for \(\Delta\) of interest.

\section{Experiments}

\subsection{Simulation Study} 
\begin{figure}
    \centering
    \includegraphics[width=0.8\linewidth]{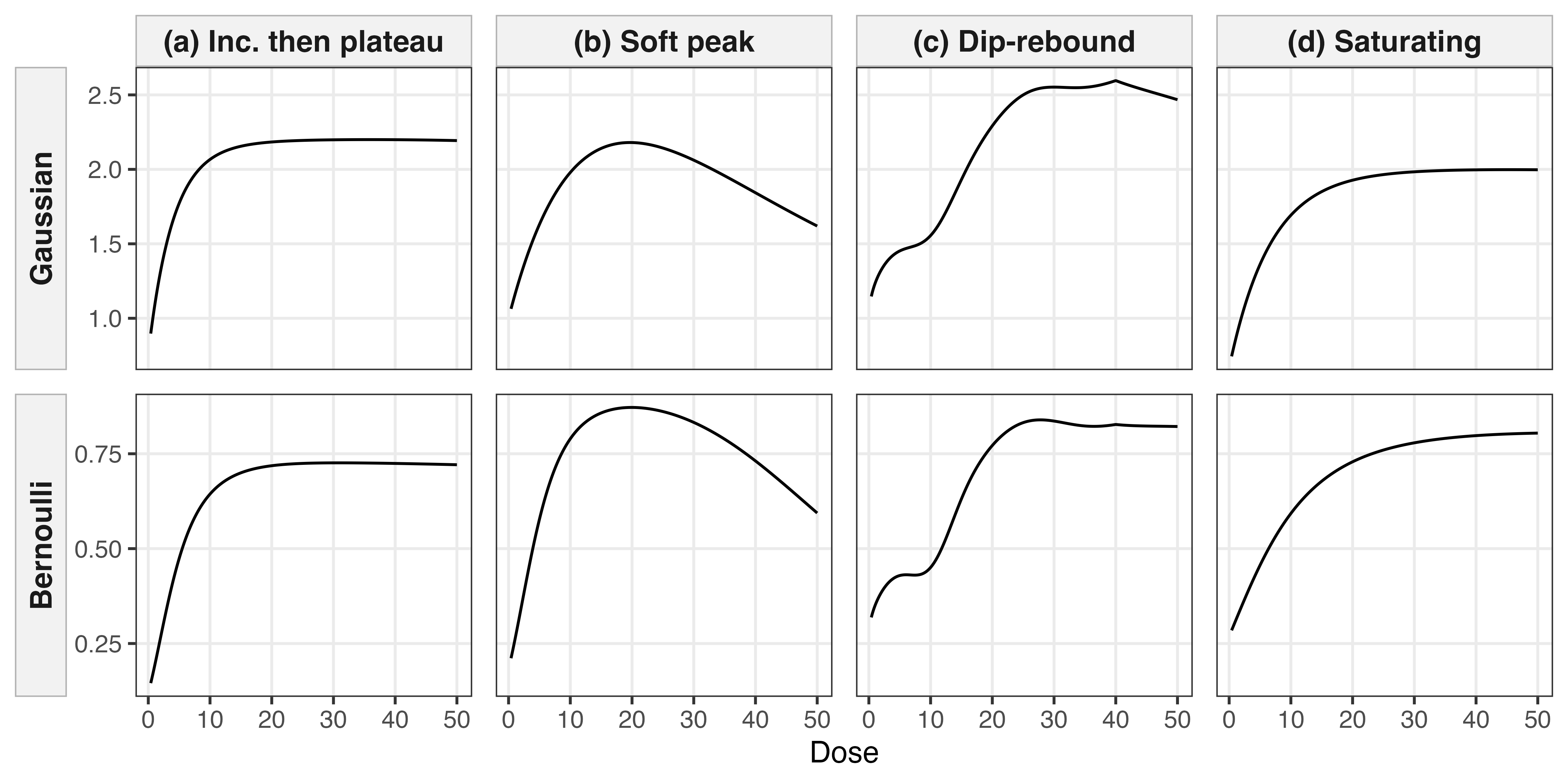}
    \caption{\small{Visualizations of Data-Generating Processes.}}
    \label{hubin:fig:dgps}
\end{figure}

We conducted a large-scale simulation study involving both continuous (regression) and binary (survival-type) outcomes. The data-generating processes (DGPs) represented typical biological patterns: (a) monotonic increase and plateau; (b) soft peaks; (c) dip-and-rebound patterns; and (d) saturating responses. See explicit functional forms for DGPs in Table~\ref{hubin:tab:dgps} visualized in Figure~\ref{hubin:fig:dgps}.  Further, \(x_i\)'s follow trials in Figure \ref{hubin:fig:real_data} with the same sample size $n$ corresponding to $8$ fishes per dose for the Gaussian outcomes and $60$ per dose for survival. Gaussian outcomes were simulated from \(\eta_i + \varepsilon_i, \varepsilon_i \sim N(0,\sigma^2),\) and \(\sigma \in \{0.1, 0.5, 1, 2, 3, 4, 5\}\), while for binary outcomes we simulated Bernoulli outcomes with \(p_i = \text{logit}^{-1}(\eta_i + \varepsilon_i)\) and $\varepsilon_i\sim N(0,0.01\cdot\sigma^2)$ to compensate for larger sample sizes.

\begin{table}[htbp]
\centering
{
\caption{\small{Data-Generating Processes (DGPs) Scenarios.}}
\label{hubin:tab:dgps}
\tiny
\begin{tabular}{lll}
\toprule
Gaussian  & Bernoulli \\
\hline
 \(\eta_a = 0.8 + 1.4 (1 - e^{-0.25 x}) - 0.0015 (x - 35)^2 / 50 \) 	&	 \(\eta_a =-2 + 3 (1 - e^{-0.2 x}) - 0.05 (x / 50)^2\) \\
\(\eta_b =1.0 + 1.2 (x / 20) e^{-0.05 (x - 20)} - 0.001 x\) 	&	 \(\eta_b =-1.5 + 4 (x / 30) e^{-0.05 (x - 25)} - 0.03 (x / 50)^2\) \\
 \(\eta_c =1.0 + 0.25 \sqrt{x} - 0.25 e^{-(x-10)^2 / 30}\) 	&	 \(\eta_c =-1 + 0.4 \sqrt{x} - 0.5 e^{-(x-10)^2 / 25}\) \\
              \(+ 0.25 e^{-(x-25)^2 / 80} - 0.03 \max(0, x-40)\) 	&	              \(+ 0.6 e^{-(x-25)^2 / 80} - 0.03 \max(0, x-40)\) \\
 \(\eta_d =0.7 + 1.3 (1 - e^{-0.15 x}) - 0.002 (x - 40)^2 / 100 \)	&	 \(\eta_d =-1 + 2.5 (1 - e^{-0.08 x}) - 0.04 (x / 50)^2\) \\
\bottomrule
\end{tabular}}
\end{table}

\begin{figure}[!h]
    \centering
    \includegraphics[width=0.8\linewidth]{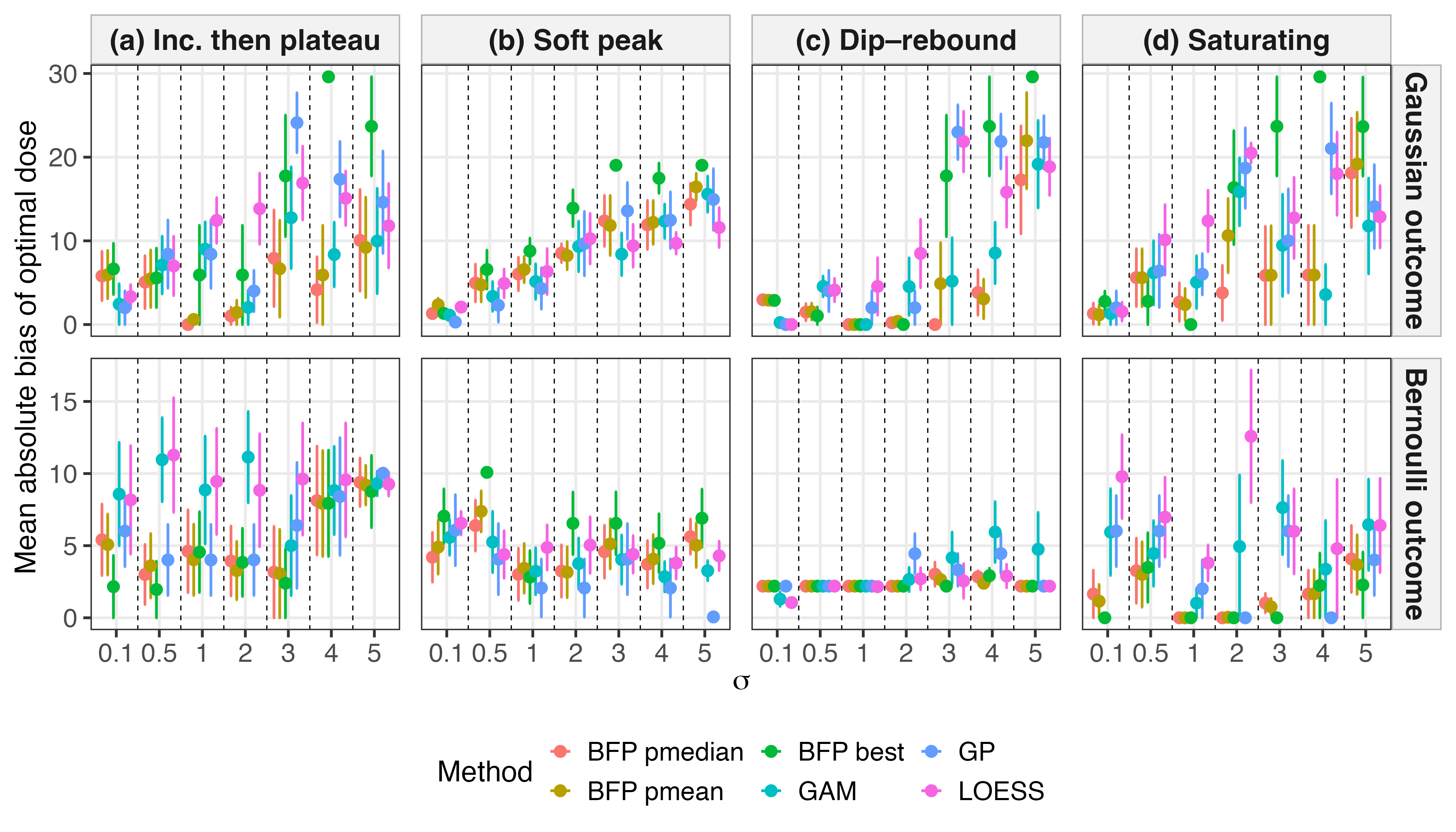}
    \caption{\small{Mean, min. and max. absolute biases per simulation setting.}}
    \label{hubin:fig:sdetailed}
\end{figure}

We compared three BFP curve prediction (for \(x\in\mathbb{R}^{[0.4,30]}\)) variants: Posterior median curve, posterior mean curve, and the curve obtained from the best model $\mathrm{m}_{\mathrm{HPM}}$ with three standard benchmark methods: GP with RBF kernel, LOESS, and GAM.  To allow for a fair comparison, all DGPs are misspecified with respect to possible closed forms solutions from the compared methods. Further, all methods including BFP used their default parameter specifications. Performance was assessed by the absolute bias in estimating the optimal dose levels (curves' maxima) across 5 replicates per noise level. Overall results are shown in Table~\ref{hubin:tab:ranking_results}, while detailed per simulation setting summaries are presented in Figure~\ref{hubin:fig:sdetailed}.

In general as expected, the methods perform generally worse with increased noise level. Further, across most settings, the BFP-based posterior mean and median curves achieved the smallest absolute biases, resulting in the lowest sum or ranks across as reported in Table~\ref{hubin:tab:ranking_results}. Specifically, BFP\_pmean and BFP\_pmedian ranked highest with the former having a median rank of 2.0 and a sum of ranks of 157.0, and the latter a median rank of 3.0 and a sum of ranks of  154.5. In contrast, GAM and GP tied at a median rank of 4.0 with sums of ranks of  204.5 and 208.5, respectively, while BFP\_best and LOESS ranked lower at 4.0 and 4.5 median ranks, with sums of ranks of  213.5 and 238.0. The sum of ranks metric is also fully consistent with the mean absolute bias ranking. 

\begin{table}[!h]
\centering
\caption{\small{Overall Ranking and Bias Summary Across Simulation Scenarios.}
\label{hubin:tab:ranking_results}}
\centering
\tiny{
\begin{tabular}[t]{lrrrrrrr}
\toprule
Method & Sum of Ranks & Median Rank & IQR Rank & Mean Abs. Bias & SD Abs. Bias\\
\midrule
BFP\_pmedian & 154.5 & 3.0 & 1.62 & 4.63 & 6.81\\
BFP\_pmean & 157.0 & 2.0 & 2.00 & 4.98 & 7.20\\
GAM & 204.5 & 4.0 & 2.12 & 6.25 & 7.41\\
GP & 208.5 & 4.0 & 2.25 & 7.10 & 8.80\\
BFP\_best & 213.5 & 4.0 & 4.00 & 8.14 & 10.48\\
LOESS & 238.0 & 4.5 & 3.00 & 8.26 & 7.85\\
\bottomrule
\end{tabular}}
\end{table}

To test whether BFP\_pmedian outperforms the baselines significantly in terms of mean absolute bias we ran a linear mixed-effects model. The results in Table~\ref{hubin:tab:lmm_absbias_methods} confirm significantly lower absolute bias than all competing methods except BFP\_pmean, for which the difference is not significant. We therefore use BFP\_pmedian in the real data application.

\begin{table}[!h]
\centering
\caption{\small Linear mixed model for absolute bias, using BFP\_pmedian as the reference method. The model includes a random intercept for each problem–scenario–noise combination and a random intercept for each simulation replicate, to account for between-setting and between-replicate variability.}
\tiny
\begin{tabular}{lrrr|l}
\hline
Method / Effect      & Estimate & $t$ value & Pr($>|t|$) & RE Var.\ (SD) \\
\hline
Intercept (BFP\_pmedian) & $4.634$ & $5.70$ & $<0.001$ & setting: $21.08\ (4.59)$ \\
BFP\_pmean               & $0.344$ & $0.60$ & $0.548$  & replicate: $0.61\ (0.78)$ \\GAM                      & $1.618$ & $2.83$ & $0.005$  
 & residual: $45.73\ (6.76)$ \\
GP                       & $2.470$ & $4.32$ & $<0.001$ &  \\
BFP\_best                & $3.506$ & $6.13$ & $<0.001$&  \\
LOESS                    & $3.621$ & $6.34$ & $<0.001$ &  \\
\hline
\end{tabular}
\label{hubin:tab:lmm_absbias_methods}
\end{table}

\subsection{Estimating Fish Nutrient Requirements} 
\begin{figure}[htbp]
\centering

% Panels
\begin{minipage}[t]{0.47\linewidth}
  \centering
  {\small\colorbox[gray]{0.93}{\parbox{\linewidth}{\centering (a) Condition factor}}}\\[0.3em]
  \includegraphics[width=\linewidth,
                   trim= 1.2cm 2.8cm 1.5cm 3.3cm,clip]{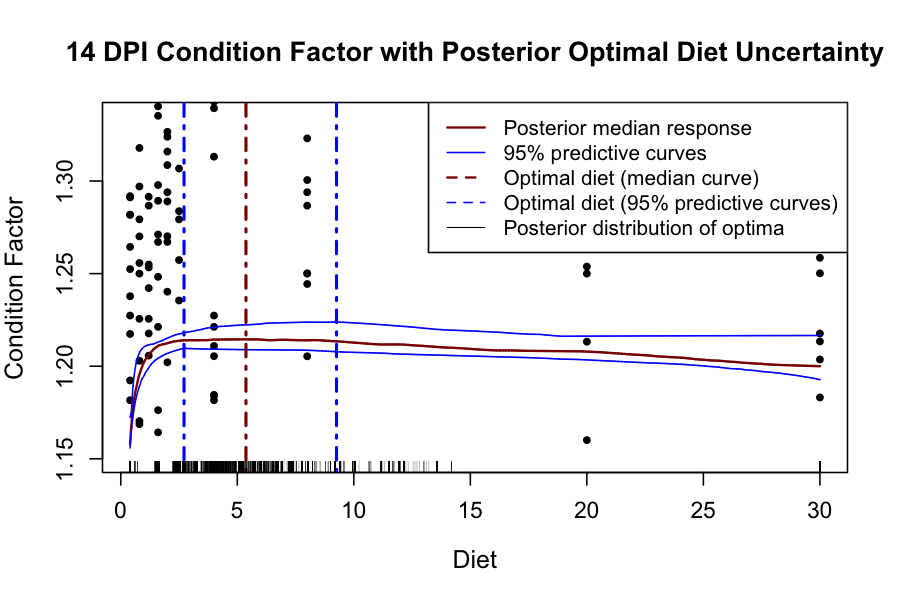}\\[0.6em]
\end{minipage}\hfill
\begin{minipage}[t]{0.47\linewidth}
  \centering
  {\small\colorbox[gray]{0.93}{\parbox{\linewidth}{\centering (b) Survival}}}\\[0.3em]
  \includegraphics[width=\linewidth,
                   trim= 1.2cm 2.8cm 1.5cm 3.3cm,clip]{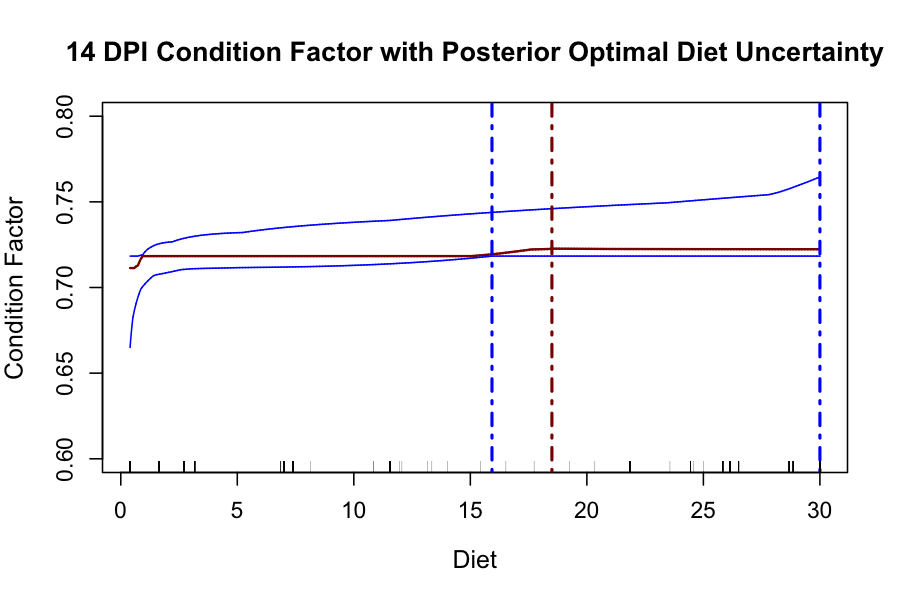}\\[0.6em]
\end{minipage}

{\small Vitamin A level (mg/kg)}

\caption{\small Dose--response curves for CF-SW (left) and survival-rates (right).}
\label{hubin:fig:real_data}
\end{figure}

The BFP framework with BFP\_pmedian predictions were further applied to an Atlantic salmon feeding and pathogen infection trial to estimate the dietary requirement and upper safe level of vitamin A. The goal was to identify the vitamin A concentration that maximizes body condition factor or survival after bacterial infection, while capturing nonlinear dose–response behavior potentially driven by nutritional saturation or toxicity at high levels. Two response variables were analyzed: condition factor after a 15-week feeding trial  in fresh water and subsequent seawater transfer (CF-SW), and survival after infection with \textit{Yersinia ruckeri}.  In each case, the BFP model was fit using \texttt{FBMS}, and posterior quantiles for the curves were obtained corresponding to medians and 95\% credible intervals (CrI). The estimated curves' optima indicated intermediate dose levels: 18.5 mg/kg for survival (95\% predictive CrI based  15.93--30 mg/kg), and 5.11 mg/kg for CF-SW (95\% predictive CrI based 2.71--9.22 mg/kg). These results align with biological expectations: Inadequate dietary vitamin A reduces growth and resilience to bacterial infection, whereas excessive levels may exert toxic or detrimental effects. The posterior mean dose–response curves, their uncertainty bands, and optima, as well as the full posterior of the optima are illustrated in Figure~\ref{hubin:fig:real_data}.

\section{Discussion and Conclusions}
The Bayesian fractional polynomial model provides a flexible framework for optimal dosage estimation. Its ability to account for both model (functional form) and parameter (regression coefficients) uncertainty within a single inference framework distinguishes it from classical smoothers or deterministic curve fitters. Through simulation, we demonstrated its robustness across (misspecified) nonlinear generative processes and multiple noise conditions, with BFP's posterior median and mean curves achieving the best overall performance, which was significantly better than non-BFP baselines.

In real biological data, the approach identified plausible optima with credible uncertainty quantification. Obtained intermediate optima in salmon diets suggest a balance between adequate vitamin A provisioning and excess, consistent with fish nutrition literature. Future research will involve informative prior specifications for the BFP basis using expert knowledge. Beyond aquaculture, the framework generalizes naturally to pharmacology, toxicology, and other dose-exposure studies.

\subsubsection*{Funding}
The biological trials of this work was supported by the Norwegian
Seafood Research Fund through the project REVITALISE (Grant 901834).

\bibliographystyle{plainnat}
\small{

}
\end{document}